\documentclass[aip,graphicx,amsmath,amssymb]{revtex4-1}

\usepackage{graphicx}
\usepackage{dcolumn}
\usepackage{bm}
\usepackage[utf8]{inputenc}
\usepackage[T1]{fontenc}
\usepackage{mathptmx}
\usepackage{etoolbox}
\usepackage{color}
\usepackage{ulem}
\usepackage{verbatim}

\makeatletter
\def\@email#1#2{%
 \endgroup
 \patchcmd{\titleblock@produce}
  {\frontmatter@RRAPformat}
  {\frontmatter@RRAPformat{\produce@RRAP{*#1\href{mailto:#2}{#2}}}\frontmatter@RRAPformat}
  {}{}
}%
\makeatother
\draft 

\begin{document}


\title[Coaxial Ion Source]{Coaxial Ion Source : pressure dependence of gas flow and field ion emission}

\author{D. Bedrane}
 \affiliation{CINaM, Aix-Marseille Univ, CNRS, UMR7325, France}
\author{A. Houël}
\author{A. Delobbe}
 \affiliation{Orsay Physics, ZA Saint-Charles, 13710 Fuveau, France}
\author{M. Lagaize}
\author{Ph. Dumas}
\author{S. Veesler}
\author{E. Salan\c{c}on}%
 \email{evelyne.salancon@univ-amu.fr}
 \homepage{https://www.cinam.univ-mrs.fr/}
\affiliation{CINaM, Aix-Marseille Univ, CNRS, UMR7325, France}

\date{\today}

\begin{abstract}
We investigated the pressure dependence of gas flow and field ion intensity of a coaxial ion source operating at room temperature over a wide pressure range, testing various gases and ionisation voltages. Flow conductance measurements taking into account the different gases’ viscosity and molecular mass consistently exhibit a generic pattern. Three different flow regimes appear with increasing upstream pressure. Since the coaxial ion source supplies the gas locally, very near the apex of the tip where ionisation occurs, large ionisation currents can be obtained without degrading the propagation conditions of the beam. Compared with field ionisation in a partial pressure chamber, using the coaxial ion source increases the ion current a hundredfold for the same residual low pressure. We also show that the gas flow regime does not impact ionisation yield. Although a fuller characterisation remains to be performed, brightness reaches $3 \times 10^{11}$ A/m$^2$/sr at 12kV extracting voltage.
\end{abstract}


\maketitle 

\section{INTRODUCTION}
The search for ion sources has a long history, marked by both stalemates and rapid progress. A range of approaches has met differing degrees of success. The development of helium ion microscopy \cite{economou2012} is a striking example of decades of cumulative patience and perseverance on the part of many. It also shows that the quality of the source itself is paramount to the overall performance of a system. Yet while this achievement is undoubtedly a milestone, it is certainly not the end of the story of bright ion sources.  Indeed, the range of possible applications implies a range of desired source properties. For Focused Ion Beam (FIB) applications, in addition to Liquid Metal Ion Sources (LMIS), Gas Field Ion Sources (GFIS) or plasma ion sources are also technical alternatives \cite{Tondare2005}. Brightness, reliability, the selection of possible ions, ease of use, sample contamination, costs, etc. are among the important parameters to be taken into account. 
Here, we report on our progress in developing our \cite{Descoins2008a,Hammadi2012,Salancon2003,Salancon2022a} GFIS. Operating at room temperature (the aim being to keep it simple) for different noble gases, it is hereafter referred to as the Coaxial Ion Source (CIS). In his 2005 review article \textit{Quest for high brightness, monochromatic noble gas ion sources}, Tondare \cite{Tondare2005} classified this source, together with Konishi, Takizawa and Tsumori's helium-cooled source \cite{konishi1988}, under the \textit{needle-in-capillary type GFIS}. As we will demonstrate, one main advantage of locally injecting the gases is that high current intensities are obtained while maintaining the pressure low enough for unimpeded beam propagation. In addition, the flow regimes of the different gases are plotted in reduced units, enabling us to reveal their generic pattern. We also describe the mapping of the field ion current under varying pressure and high voltage. Brightness is estimated, and the promising results (compared with LMIS) of this noble gas ion source operating at room temperature encourage us to pursue the development of our CIS.

\section{Materials and methods}
\label{MaterialsandMethods}
The heart of the experimental device is the coaxial ion source (CIS) \cite{Salancon2003,Descoins2008a,Hammadi2012,Salancon2022a}. It constitutes a deliberate leak between a high-pressure vacuum chamber and a low-pressure vacuum chamber. When a sufficiently high voltage is applied to the inserted needle tip, a portion of the molecules entering the low-pressure chamber through the coaxial ion source is ionised. This high voltage is controlled by a computer program which records the pressure and the tip emission high voltage and current.

\begin{figure}
\includegraphics{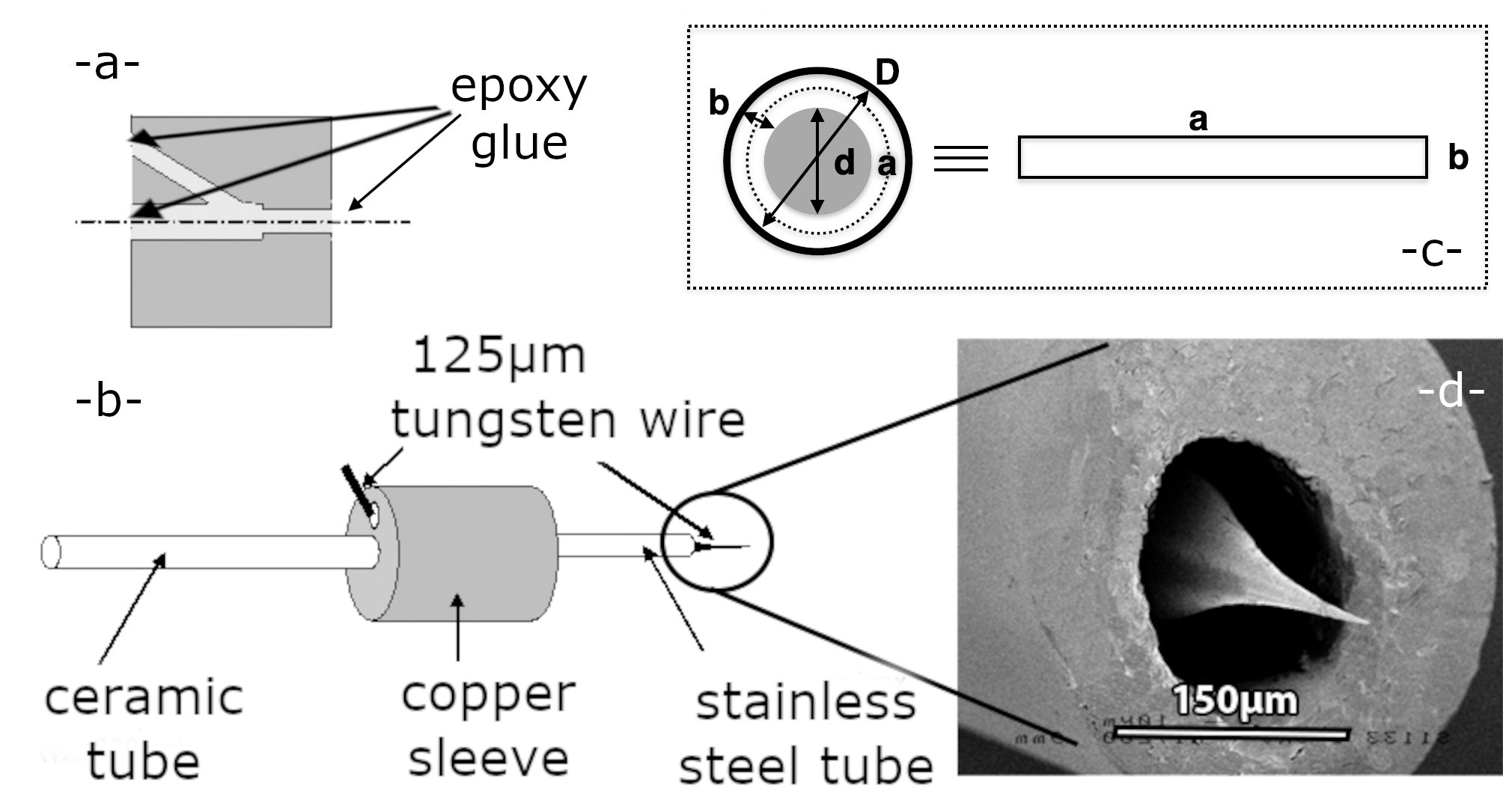}
\caption{Schematic of coaxial ion source: -a- Cross-sectional view of copper sleeve connecting the ceramic tube (left) and the stainless steel tube (right) into which the tungsten wire is inserted; -b- General view of CIS structure; -c- Geometric modelling parameters of CIS cross-section; -d- Scanning electron microscopy image of CIS low-pressure outlet.
\label{coax}}
\end{figure}

\subsection{Coaxial ion source}
\label{coax_sec}
The coaxial ion source consists of a stainless-steel capillary tube guiding a tungsten wire shaped into a nanotip, as presented in Figure \ref{coax}. A tungsten wire (diameter: $d = 125 \mu$m) is first inserted into a stainless steel tube (length $L=(6,0\pm0,2)$mm, diameter: $D = 170 \mu$m) maintained in a copper sleeve. The tip is prepared by electrochemical etching \cite{Muller}; then, to minimise corrosion \cite{Hammadi2012,Anway1969}, palladium is electrochemically deposited \cite{Kuo2004,Kuo2006,Nakagawa2013}. The degree to which the apex of the tungsten tip protrudes from the capillary tube (see Figure \ref{coax}-d-) is controlled by retracting the wire. The length of tip chosen here to emerge from the tube is comparable with the wire diameter $ \approx130\mu$m. The stainless-steel capillary tube and the tungsten wire are appropriately glued to the copper sleeve (see Figure \ref{coax}-a-) with epoxy, to ensure sealing while preserving gas flow along the CIS. Epoxy is also used to glue the ceramic tube that insulates the CIS from the electrical ground and conveys the gas that will be injected coaxially near the apex of the tip from the high-pressure chamber (see Figure \ref{coax}-b-).

\subsection{Vacuum chambers}
\label{pump}
\begin{figure}
\includegraphics{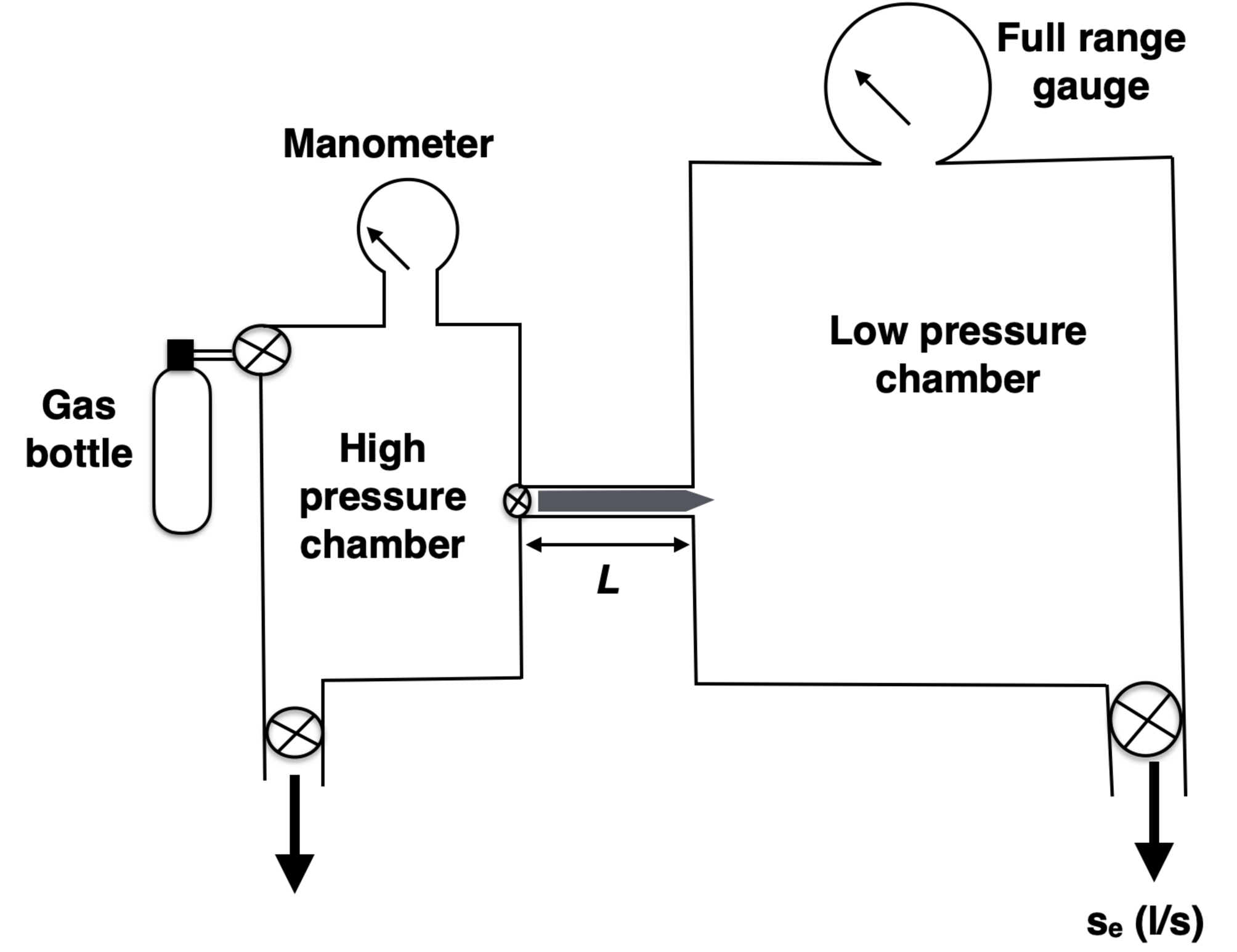}
\caption{Experimental setup: The coaxial ion source, of length L, separates a high-pressure chamber from a low-pressure chamber. \label{exp}}%
\end{figure}

As described above, the CIS is a deliberate leak between a high-pressure chamber and a low-pressure chamber (see Figure \ref{exp}). The high-pressure and low-pressure vacuum chambers can independently be evacuated down to, respectively, high or ultra-high vacuum conditions after baking, via a turbo-molecular pump (TwisTorr305, Agilent). As we will see below, to calculate the molecular conductance of the CIS, we need to know the pumping speed. While intrinsic pumping speeds are provided by the manufacturer of the turbo-molecular pump, the effective pumping speeds of our experimental setup will be measured (see Section \ref{Results} and Appendix \ref{effpumpsped}). 
\\ Gases come from Linde minican bottles and are up to $99.5\%$ pure. A leak valve provides the gas inlet from the bottle in the high-pressure chamber. For our experiments, the high-pressure $P_{hp}$ is maintained between $10^{2}<P_{hp}<10^{5}$ Pa. In a typical experiment, $P_{hp}$ is set at the desired value, while the low-pressure chamber alone is evacuated to balance the incoming gas. This creates a low-pressure equilibrium $P_{lp}$ in the low-pressure chamber.
The pressure in the high-pressure chamber is measured with a membrane-based gauge (ED 510/421.411/025 from Haenni). A full-range gauge from Varian is used in the low-pressure chamber; a gas-dependent gauge factor is required. The gauge factors are given in Table \ref{speed}. The low-pressure chamber is also equipped with a quadrupole mass spectrometer (QMG-064 from Balzers) whose bandwidth reaches $20$Hz.
In the low-pressure chamber, the CIS is mounted on a rotary manipulator allowing the emission angle of the beam to be measured (see Appendix \ref{emissionangle}). For flexibility, a hollow stainless steel coil connects the CIS and the high-pressure chamber. The overall flow conductance of the deliberate leak is limited by the flow conductance of the CIS itself.
The low-pressure chamber is also equipped with an adjustable extraction grid, a Faraday cup and a multichannel plate/fluorescent screen assembly used for projection microscopy \cite{Salancon2022a,Salancon2003}.

\subsection{Electrical measurements}
The low-pressure UHV chamber is equipped with several high-voltage feedthroughs connected to the CIS, the extraction grid, the Faraday cup, the electrodes of the multichannel plate and the fluorescent screen.
\\For most experiments, the CIS is virtually grounded through a transconductance amplifier (Model 427, Keithley) which provides the total ionisation current from the tip. The high voltage for extraction is supplied to an extraction grid (electron microscopy copper grid: square pattern 150 Mesh TEM Support Grids from Agar Scientific) about $5$mm away from the apex of the tip. The power supply is a high-voltage 30kV, reversible, low-ripple module (MPS30 from Spellman).
\\For stabilised pressure conditions, the high voltage is ramped up and down like a staircase. The voltage steps are computer-controlled (see Section \ref{curmeas}). Each step up corresponds to $1$kV and lasts $t\approx 5sec$. The rise time of the Keithley427 is set at $0.01$s and the output is sampled by the acquisition program and averaged (see Section \ref{curmeas}).
The control and acquisition program is written in Labview2011 version 11.0.1 interfacing a National Instrument analog and digital input/output card (USB-6009). After acquisition, data are processed and plotted using Igor Pro 9.
\subsection{Beam characterisation}
While this article details the measurement of CIS ion intensity over a wide range of pressures and high voltages, the full 2D-mapping of brightness when pressure and high voltage vary is beyond its scope. However, although brightness is of the utmost importance as far as source characterisation is concerned, it can be estimated from measurements in certain configurations. Determining brightness also requires knowing the size of the virtual source and the emission solid angle. The size of the virtual source is derived from a projection microscopy image (see Appendix \ref{emissionangle}), using a procedure described elsewhere \cite{Salancon2022a}. Briefly, we project a mask, such as a lacey carbon film, onto the fluorescent screen of an opaque, hollowed, thin object. If the source is spatially extended, despite the rapidly varying transmission contrast of the mask, contrast variations in projected image intensity will be blurred. Based on these intensity profiles and the degree of magnification, an upper limit to the size of the virtual source can be obtained. The fluorescent screen, $4$ cm in diameter and $40$ cm away from the tip, receives only part of the ion beam. To estimate the emission angle, the intensity projected on the fluorescent screen was observed while tilting the rotary manipulator mentioned above.
\section{Results}
\subsection{Pressure measurements}
\begin{figure}
\includegraphics{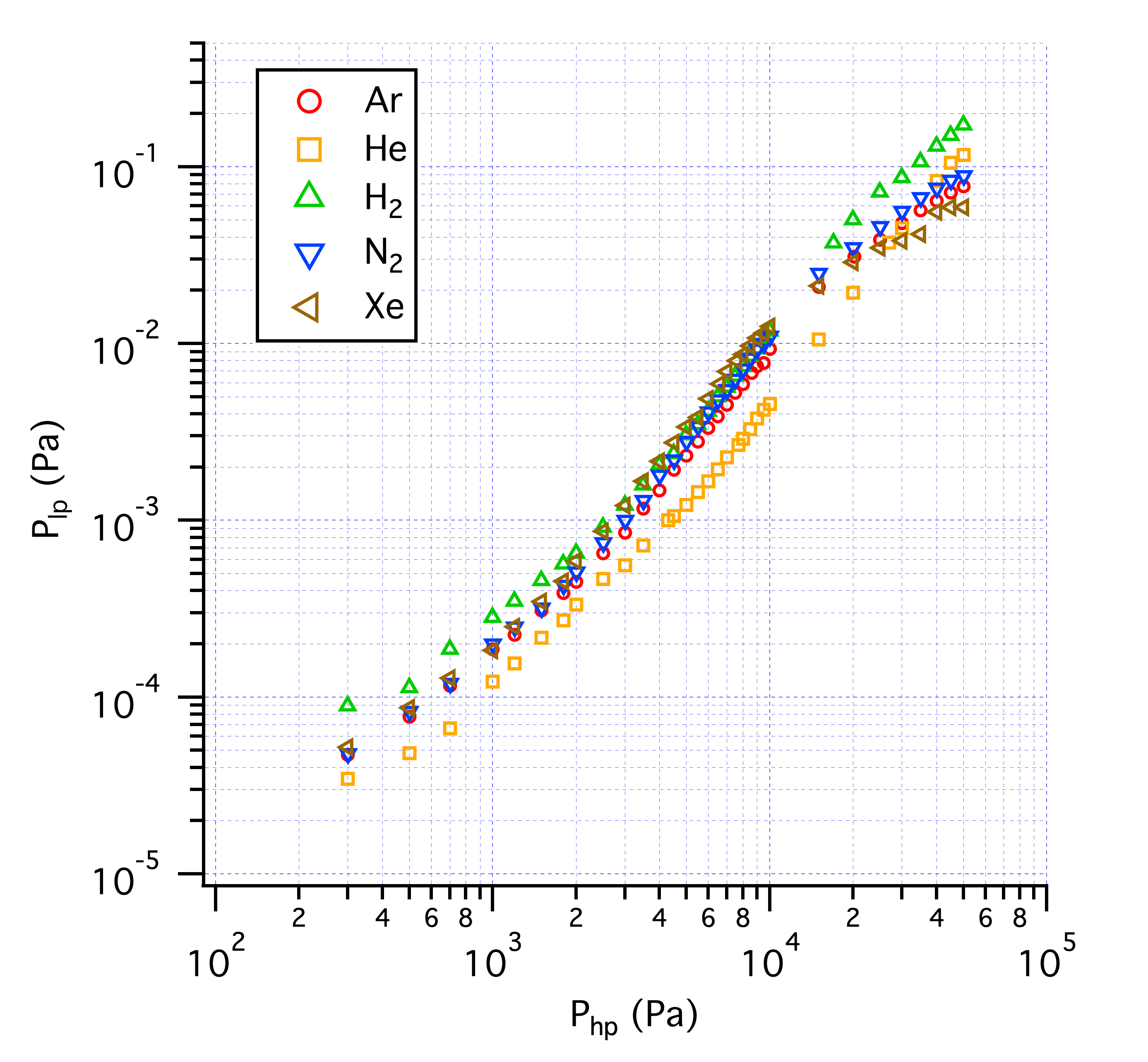}
\caption{\label{Pressure} Pressure in the low-pressure chamber $P_{lp}$ vs. pressure in the high-pressure chamber $P_{hp}$ for 5 different gases.}
\end{figure}

In a log-log plot, Figure \ref{Pressure} displays the $P_{lp}$ for $5$ different gases when the $P_{hp}$ is set within a three-decade range. $P_{lp}$ increases with $P_{hp}$. We observe that the order of magnitude of $P_{lp}$ / $P_{hp} \approx 10^{-6}$, resulting in pressures in the low-pressure chamber compatible with unimpeded ion propagation.
\subsection{Effective pumping speed measurements}
\label{Results}
 The effective pumping speed \cite{steckelmacher1966} $S_{e}$, needs to be known in order to plot the conductance. It is measured by monitoring the (exponential) decay of $P_{lp}$ after closing the valve between the high- and low-pressure chambers, upstream of the CIS itself. Straightforwardly, neglecting the final limit pressure in the low-pressure chamber: $$P_{lp}=P_{lp0}\exp(-\frac{S_{e}}{V}t)$$ (where $P_{lp0}$ is the initial pressure and $V$ is the volume of the low-pressure chamber). Starting from typical $P_{lp0}=10^{-2}$Pa, $P_{lp}$ decreases in a few seconds, which requires a sufficient bandwidth for pressure measurements. Based on these pressure measurements, performed with the quadrupole mass spectrometer, $S_{e}$ is calculated for four different gases from the exponential fit of these data (see Appendix \ref{effpumpsped}). The results are summarised in Table \ref{speed}.
 
\begin{table}
\caption{\label{speed} Effective pumping speeds and gauge factors for the gases. Effective pumping speeds measured as explained in Section \ref{Results}, except for Xe: being too heavy. Xe effective pumping speed is determined from reduced conductance curves (see Figure \ref{Conduc}-b)}
\begin{ruledtabular}
\begin{tabular}{c|c|c|c|c|c}
gas & N$_2$ & He & H$_2$ & Ar & Xe\\
\hline
pumping speed ($\times 10^{-3}$m$^3$.s$^{-1}$) & 98 & 206 & 177 & 88 & (65)\\
\hline
gauge factor & 1 & 0.18 & 0.46 & 1.29 & 2.87\\
\end{tabular}
\end{ruledtabular}
\end{table}

\subsection{Current measurements}
\label{curmeas}
\begin{figure}
\includegraphics{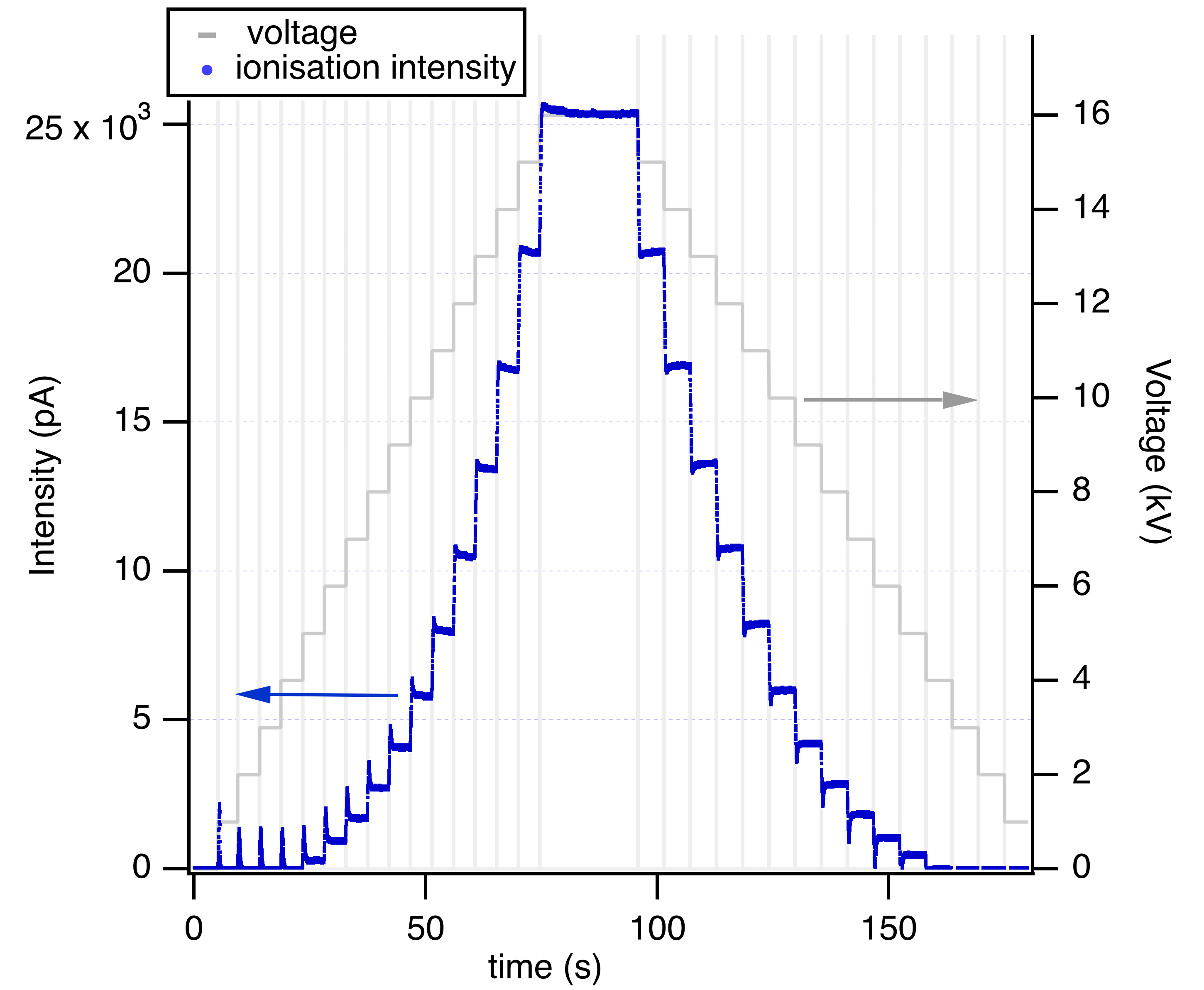}
\caption{Intensity measurement (blue, left) and applied voltage (grey, right) vs. time during a triangular, staircase voltage ramp. Each voltage step corresponds to $\Delta V=1$kV and lasts 5 seconds, except for the longer plateau at maximum voltage (16kV).
\label{It}}
\end{figure}

For selected and stabilised pressure conditions, $I(V)$ curves can be recorded. Figure \ref{It} shows both the current $I$ and the voltage $V$ as a function of time $t$ during a $3$-minute scan for a CIS using Argon in the high-pressure chamber ($P_{hp}=3000$ Pa). It can be seen that, as expected for a field ionisation process, current increases more rapidly than voltage. The current spikes at the beginning of each plateau are only due to capacitive coupling of the setup, and are taken into account by the software for the post-processing procedures. A stable source being required throughout the entire experiment, the current is measured during a voltage round trip to check that the apex of the tip has not evolved due to the high-voltage conditions.

\section{Discussion}
\subsection {Gas flow regimes}
\label{conductance}
\begin{figure*}
\includegraphics{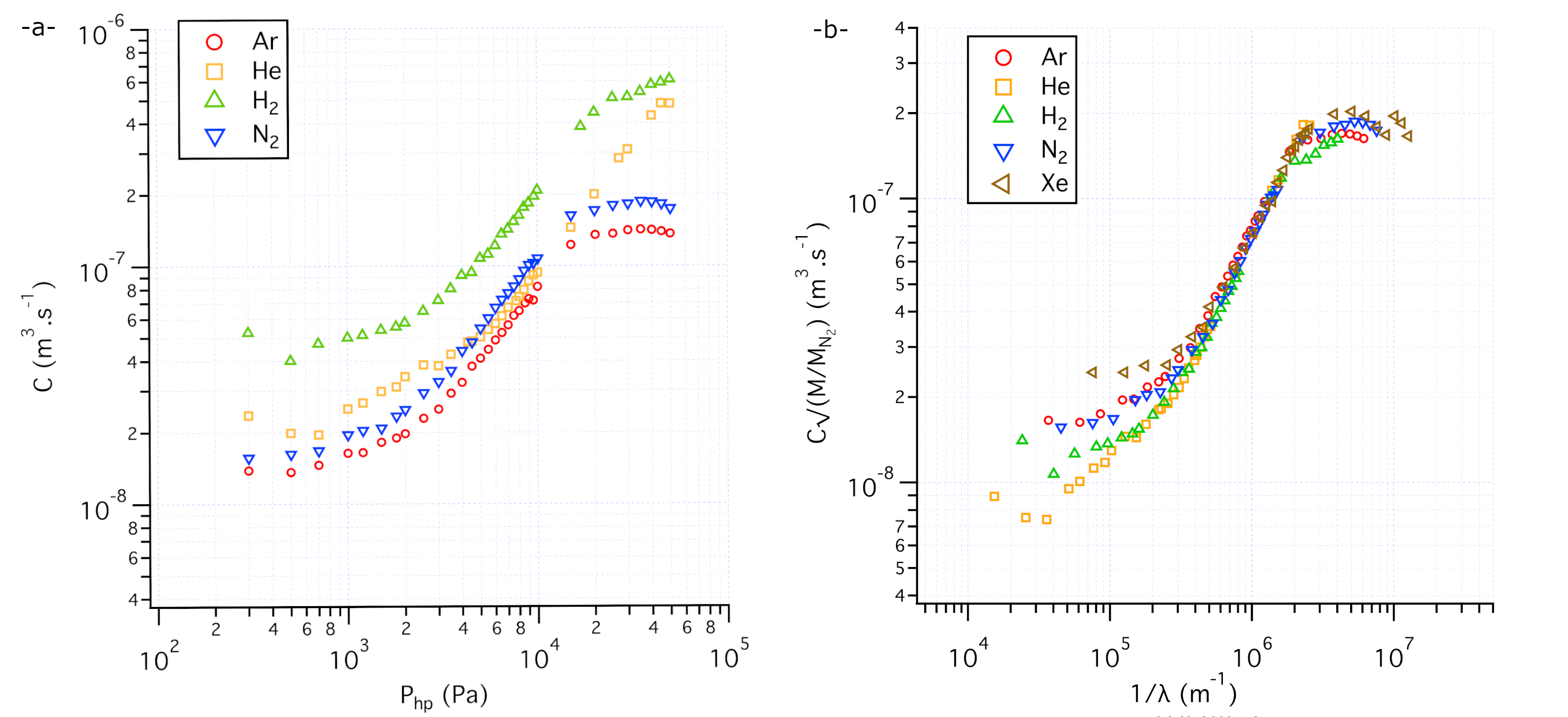}
\caption{\label{Conduc} -a- Conductance vs. pressure in the high-pressure chamber; -b- Reduced conductance $C^{*}$ vs. $1/\lambda$, the inverse of the mean free path of the high-pressure chamber for 5 gases.}
\end{figure*}

Gas flow regimes are described in the literature \cite{vacuum,Descoins2008a,hanlon} through conductance $C$, defined as the ratio of molecular flux $Q$ to the difference in pressure $(P_{hp}-P_{lp})$ : $$C= \frac{Q}{P_{hp}-P_{lp}} \approx \frac{S_{e}P_{lp}}{P_{hp}}$$
where $Q$ is the product of the effective pumping speed $S_{e}$ by the low pressure and where we consider $P_{hp}>>P_{lp}$. Figure \ref{Conduc}-a- shows the conductance of a CIS structure plotted versus the pressure in the high-pressure chamber. For each of the four gases plotted, the figure exhibits an S-shaped curve. Two plateaus, separated by a regime where the conductance $C$ increases with $P_{hp}$, can be seen. This behaviour points to a succession of three different gas flow regimes as $P_{hp}$ increases:
\begin{itemize}
    \item At low pressure, the mean free path being larger than the characteristic length scale of the CIS structure, the gas flows in a molecular regime. This is the first conductance plateau. 
    \item When pressure increases, viscosity starts to play a dominant role and the gas flow is now in a viscous (or laminar) regime. In this regime (here $10^3 \lesssim P_{hp} \lesssim 10^4$ Pa), conductance $C$ increases linearly with $P_{hp}$.
    \item Finally, conductance is limited by the cross-section of the CIS structure to what is known as a choked-flow regime, and reaches the second plateau. These results refine and extend the findings and description previously reported by Descoins et al. \cite{Descoins2008a}.
\end{itemize}
Numerous authors \cite{Knudsen,steckelmacher1966,Steckelmacher1986,vacuum, hanlon} have derived analytical and/or semi-empirical expressions of conductance $C$ in various flow regimes and channel geometries, from the simplest cylindrical tube to more complex channel geometries. In these expressions, conductance appears as a separable function. When an analytical expression can be written, for each flow regime, the conductance $C_{regime}$ can be written as a product of two functions. One, $C_{regime,geometry}$,  depends only on the geometrical parameters and the other, $C_{regime,gas}$, depends only on the gas parameters: $$C_{regime}= \left[C_{regime,geometry}\right] \times \left[C_{regime,gas}\right]$$
As suggested by Descoins et al. \cite{Descoins2008a}, flow behaviour through a coaxial CIS structure can be described using a rectangular aperture of area $a\times b$  (where $a$ is the perimeter of the annular aperture between the tungsten wire and the stainless steel capillary and $b<<a$ is the width of this annular aperture).
In this rectangular geometry, the expressions of $\left[C_{regime,geometry}\right] \times \left[C_{regime,gas}\right]$ for the different regimes are:
\begin{itemize}
    \item molecular-flow regime:
    \begin{equation}
        \left[\frac{ab^2}{8L} (1+2\ln\frac{2a}{b})\right] \times \left[\sqrt{\frac{8kT}{\pi m}}\right] \label{eq:mol}
    \end{equation}
    \item viscous-flow regime:
    \begin{equation}
        \left[\frac{\pi ab^3}{96L}\right] \times \left[\frac{1}{\lambda} \sqrt{\frac{8kT}{\pi m}}\right] \label{eq:visc}
    \end{equation}
    \item choked-flow regime:
    \begin{equation}
        \left[C_{0}\sqrt{\frac{\pi}{8}} ab \right]\times \left[\Gamma \sqrt{\frac{8kT}{\pi m}}\right] \label{eq:chock}
    \end{equation}
\end{itemize}

In the above equations, the first term of the product within brackets is $C_{regime,geometry}$ while the second term of the product within brackets is $C_{regime,gas}$. $\sqrt{\frac{8kT}{\pi m}}$ is the mean velocity of the kinetic theory of gases. $\lambda$ is the mean free path of gas molecules in the high-pressure chamber, which depends on the pressure and on the viscosity $\eta$ of the gas. $\Gamma$ also depends on the gas parameters alone, being $0.725$ for monoatomic gases and $0.685$ for diatomic gases \cite{vacuum}.
Finally, the orifice coefficient $C_{0}$ is a geometry-dependent, empirical correction not expected to vary significantly for our different CIS structures. Its order of magnitude is $0.1$ \cite{Descoins2008a,santeler1986}.
The separability of $C$ as a product of functions depending either on geometrical parameters alone or on gas parameters alone will allow us to plot the same data to directly compare different gases. While all the experiments are performed at room temperature, the above analytical expressions call for us to plot (see Figure \ref{Conduc}-b-), $C^{*}= C\sqrt{\frac{m}{m_{N_2}}}$ versus $\frac{1}{\lambda}$ rather than $C$ versus $P_{hp}$ (as previously done in Figure \ref{Conduc}-a-). Our choice for the expression of the reduced conductance $C^{*}$ for the vertical axis is based on the $1/\sqrt{m}$ dependence of the conductance for the molecular- and choked-flow regimes. Nitrogen being commonly studied \cite{Hanlon1987}, for ease of comparison with the literature, we have also normalised by $m_{N_2}$. $C^{*}$ and $C$ are thus in the same units in m$^3$.s$^{-1}$. For the horizontal axis, guided by the expression of the conductance in the viscous-flow regime, taking into account the expression of $C^{*}$, the appropriate parameter becomes the inverse of the mean free path $1/\lambda=(P_{hp}/\eta)\sqrt{2m/\pi kT}$.
The S-shaped curves plotted with these axes can be seen in Figure \ref{Conduc}-b-. They are now all grouped in a single, generic curve, showing that most of the underlying physics has been taken into account. While deliberately not plotted in Figure \ref{Conduc}-a-, Xe data are now plotted in Figure \ref{Conduc}-b-. Indeed, the effective pumping speed $S_{e}$, needed to calculate $C$, proved impossible to determine using the mass spectrometer, due to Xe’s high mass. $S_{e}$ for Xe, reported in Table \ref{speed}, was not measured but arbitrarily chosen to superimpose the Xe S-shaped curve on the generic curve.\\
Our results remain quantitatively consistent when the geometrical parameters of the CIS are varied: an example is given in Appendix \ref{CvsGeom}, where similar behaviour is observed. All S-shaped curves are also grouped in another, geometry-dependent generic curve. Although this consistency reflects that most of the underlying physics has been taken into account, we are aware that the reality is more complex across the 6mm-long CIS structure: different flow regimes coexist. The discrepancies in the generic curve that can be observed in Figure \ref{Conduc}-b- probably testify to these less obvious behaviours. Moreover, the tungsten wire and the stainless steel capillary, while likely to be co-linear, may not be coaxial. A deeper analysis of these issues is beyond the scope of this article; our target is the first-order pressure dependencies of gas flow and field ion emission, which we will now discuss.

\subsection{Field ion emission}
\subsubsection{Current versus voltage}
\begin{figure}
\includegraphics{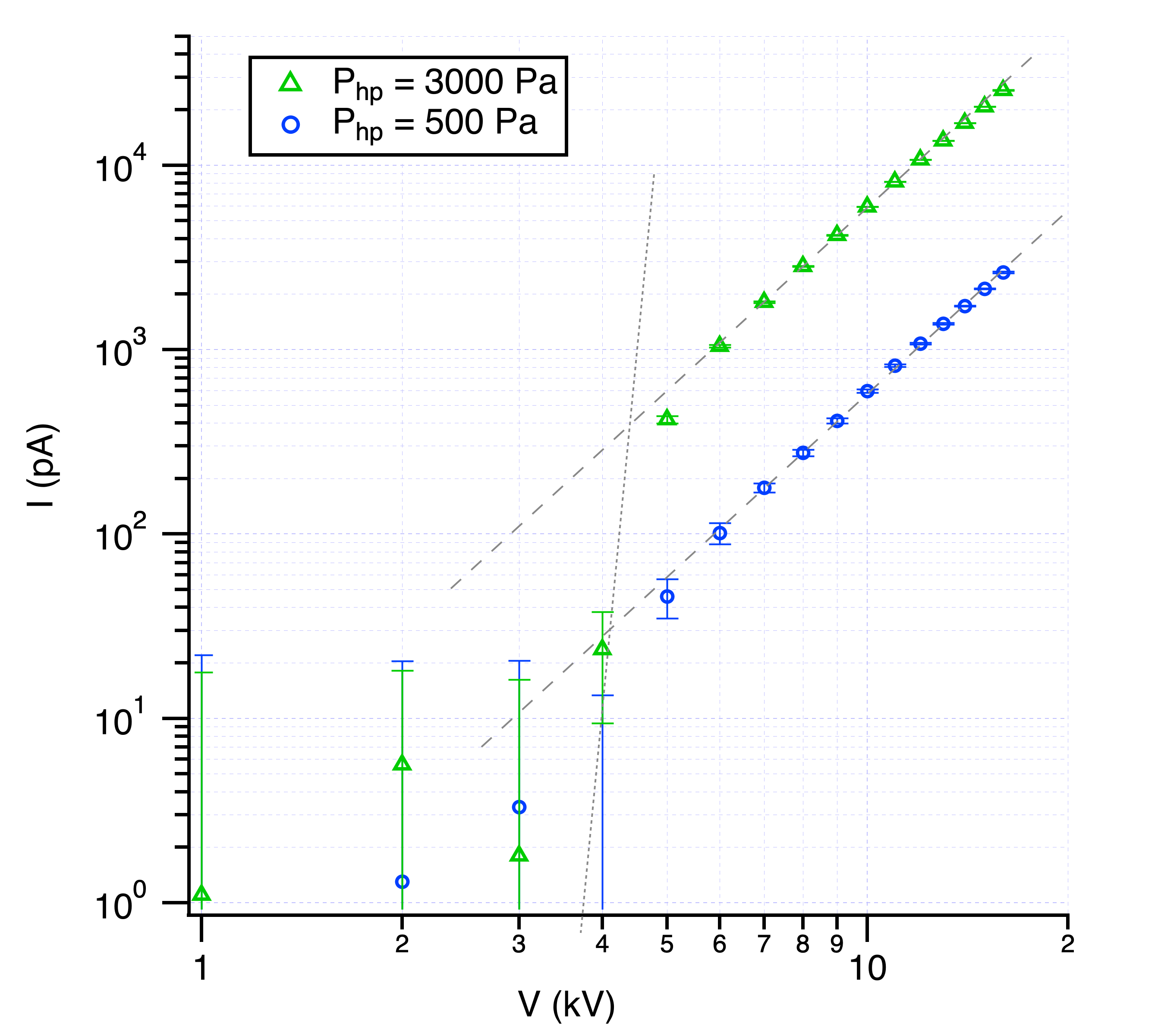}
\caption{\label{IV} Ionisation intensity vs. extraction voltage for two different pressures: $P_{hp}=500$ Pa (blue) and $P_{hp}=3000$ Pa (green). Corresponding pressures in the low-pressure chamber are respectively $P_{lp}=1.2\times10^{-4}$ Pa (blue) and $P_{lp}=1.3\times10^{-3}$ Pa (green).}
\end{figure}

From intensity data recorded versus time, like those plotted in Figure \ref{It}, we average the values obtained on the two plateaus for the same voltage while increasing and decreasing the voltage. The capacitive component of the current being opposite for increasing and decreasing voltage steps, this procedure cancels the capacitive current and gives the field ion emission current alone. Figure \ref{IV} plots in log-log scales current intensity measured versus applied voltage for Argon at two different pressures. Two radically different regimes can be seen, as observed since the early days of field ionisation \cite{Muller1955} under partial pressure conditions.
\begin{itemize}
    \item In the low-voltage regime, below $\approx 5$kV, ionisation is limited by the electric field. A slight increase in applied voltage, and thus in electric field, considerably enhances the likelihood of ionisation, thereby increasing the current. When plotted in log-log scales, a phenomenological adjustment by a straight line results in huge typical slopes of the order of several dozens \cite{Muller1955,Gomer,jousten1988}.
    \item In the high-voltage regime, above $\approx 5$kV, the ionisation probability of a molecule near the apex reaches 100 $\%$. The ionisation rate is limited by gas supply, which however also increases with voltage, as observed through the current increase. Straight-line adjustment of this regime results in typical slopes of the order of a few units. In the case of Figure \ref{IV}, the slope is $\alpha \approx 3.3$.
\end{itemize}
We also note, in Figure \ref{IV}, that the two curves are shifted by a decade, corresponding to the ratio of the two pressures $P_{lp}$ chosen for display.

\subsubsection{Current versus pressure}
Figure \ref{I_P} shows the patterns of dependence of intensity (left vertical axis) on pressure in the low-pressure chamber for three different experiments. Dashed lines are a guide for the eyes, with a slope $=1$. The linear dependence of the current with respect to pressure in the low-pressure chamber $P_{lp}$ can be observed.
The green and black current curves, indicating the highest currents, correspond to measurements from the CIS geometry at two different voltages, respectively $7.0$ kV and $12.0$ kV.
Based on these data, we can write the empirical dependence of the current $I$ versus measured pressure in the low-pressure chamber $P_{lp}$ and applied voltage $V$. 
$$I = K \times P_{lp}V^\alpha$$
where $K \approx 1000$ pA/Pa/kV$^\alpha$ and $\alpha \approx 3.3$ when $I$, $P_{lp}$ and $V$ are respectively in pA, Pa and kV, with $K$ also dependent on the CIS structure under study. Notably, this reflects the efficiency of bringing the injected gas into the ionisation zone.

\subsubsection{Ionisation yield}
\begin{figure}
\includegraphics{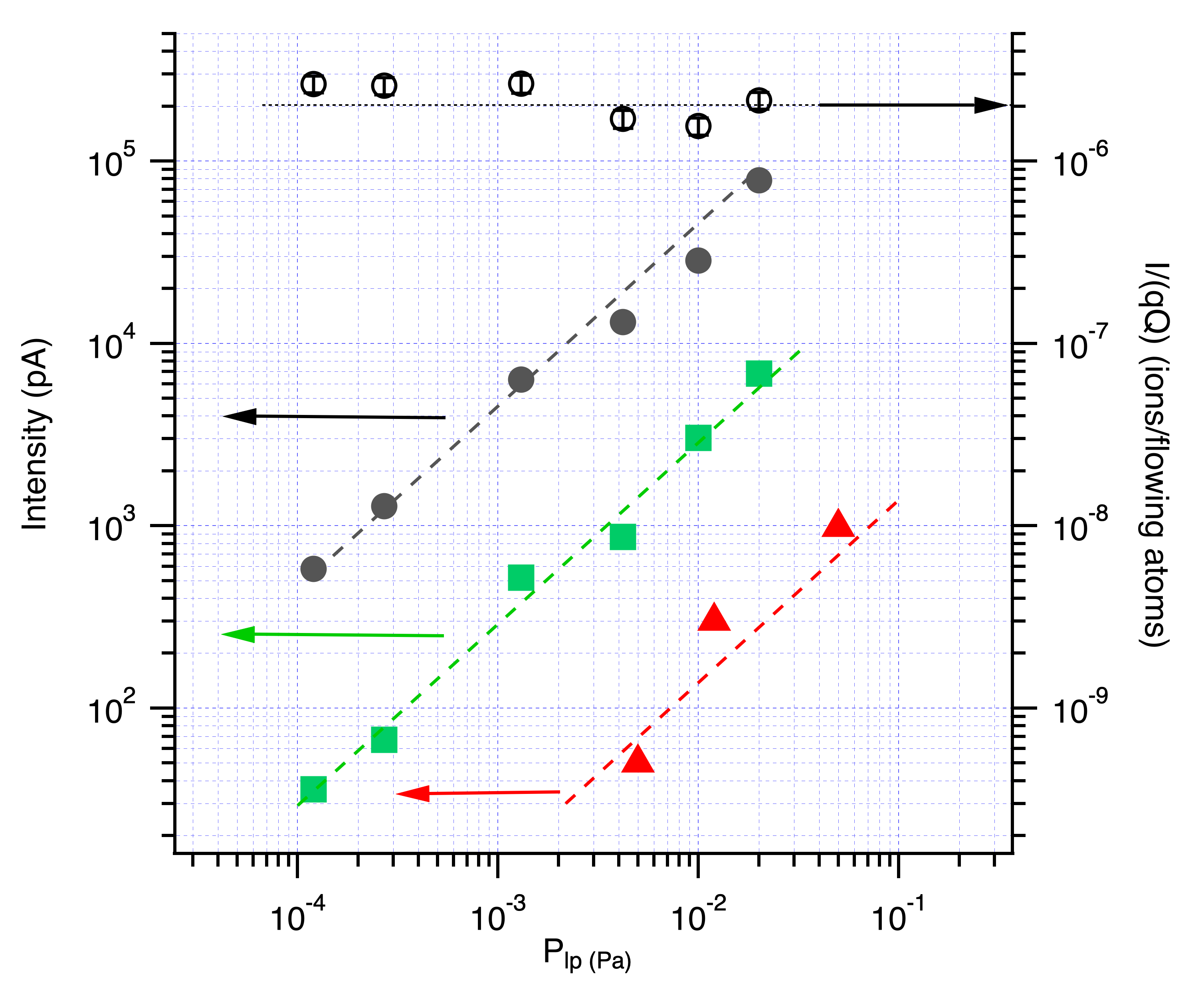}
\caption{\label{I_P} Intensity $I$ (vertical left axis) vs. $P_{lp}$ for three different experiments: for a given CIS geometry, at $V=12.0$ kV (solid black circles) and at $V=7.0$ kV (solid green squares) and, finally, for a partial pressure experiment at $V=12.2$ kV (solid red triangles). The vertical right axis shows the ionisation yield $I/(qQ)$ for the CIS geometry at $V=12.0$ kV only (open black circles). Unity slope dashed lines are a guide for the eyes. Horizontal dotted line shows that $I/(qQ)$ is almost constant.}
\end{figure}

We can estimate the ionisation yield, which is the ratio of ion flux to molecular flux $Q$, using the gas flow analysis described above. $I/(qQ)$ ($q$ is the elementary charge) is the right vertical axis of Figure \ref{I_P}. As expected from our observations, $I/(qQ)$ is almost constant over three decades in pressure. The order of magnitude of this ratio is $2\times 10^{-6}$ for $12.0$ kV. That means there are two Ar ions for one million Ar molecules entering the chamber through the CIS structure. This value is both large and improvable, as will be seen later.

\subsubsection{Advantages of the CIS}
The third curve (red triangles) in Figure \ref{I_P} is a current measurement performed with a similar tip in a partial pressure experiment. In such experiments, gas is provided through a static partial pressure $P_{lp}$, rather than dynamically provided through the CIS structure. For the same pressure $P_{lp}$, the observed currents are orders of magnitude lower than those obtained when gas is injected through the CIS structure. For $P_{lp}=10^{-2}$ Pa, read current values for the black (CIS structure) and red (partial pressure) dashed lines are respectively 45000pA and 150pA, resulting in a current ratio of ca. 300. This reflects the fact that the pressure $P_{tip}$ in the vicinity of the tip is ca. 300 times higher than $P_{lp}$. By injecting gases locally, the CIS geometry thus allows much higher ion currents to be obtained while preserving the vacuum in the low-pressure chamber. This major enhancement, which could be improved if needed, is supported by the literature. Also using a setup allowing He to be injected in the vicinity of the tip, Konishi et al. \cite{konishi1988} reported an enhancement of the same order of magnitude: $P_{tip}/P_{lp}\approx$ 110. More indirectly, from experiments performed in the partial pressure configuration, Jousten et al. \cite{jousten1988} derived a semi-empirical expression of ionisation current versus pressure. According to their semi-empirical expression, to obtain the same ionisation current as that afforded by the CIS geometry would require a pressure in the low-pressure chamber a hundredfold higher. Although slightly quantitatively different, these findings confirm the advantage of using the CIS geometry. A gain of at least two orders of magnitude is obtained when locally injecting the gas with a CIS structure of this geometry.\\
As mentioned above, the ionisation yield is $I/(qQ) \approx 2\times 10^{-6}$ ion/molecule for $12.0$ kV. This value is both large and improvable: the advantages of the CIS geometry allow a hundredfold increase in ionisation current $I$ for the same $P_{lp}$ and, if needed, the yield could be improved using other geometries. Most of the molecules are too far from the tip apex for ionisation. Indeed, the yield reflects the ratio $\approx (100nm/100\mu m)^2$ of the area capturing electrons to the aperture of the stainless steel capillary. Appropriate engineering solutions would probably increase both ionisation yield and local pressure enhancement.

\subsubsection{Brightness}
Like previous authors \cite{Borries1939,Sciaini2019, Sudraud1978}, we define brightness $B=I/(\Omega.s_{source})$, where $\Omega$ is the opening solid angle of the emitted ion beam and $s_{source}$ is the size of the virtual source. As mentioned in Section \ref{MaterialsandMethods}, to obtain the brightness $B$ for selected pressure and high-voltage conditions, we measured the solid angle and the emission area from projection microscopy experiments. At $V=12$ kV and $P_{lp}=2\times 10^{-2}$ Pa, a $\approx 5^{\circ}$ half opening angle of the emitted beam resulting in $\Omega=0.024$ sr. (see Appendix \ref{emissionangle}). From a shadow experiment \cite{Salancon2022a}, we also estimated the emission area of the virtual source $s_{source}\approx \pi(2nm)^2$. This value is probably overestimated due to mechanical vibrations from our experimental setup. We hence obtain a brightness $B\approx 3 \times 10^{11}$ A/m$^2$/sr. Although the CIS is still in development, its room-temperature brightness compares favourably with the brightness of well-established Liquid Metal Ion Sources (LMIS). Typically operated at extraction voltages $V\approx30$ kV, LMIS sources exhibit typical brightnesses \cite{Mair1996, Tondare2005} of $B\approx 3 \times 10^{10}$ A/m$^2$/sr, one order of magnitude lower. However, while we studied the dependence of current on high voltages (see Figure \ref{IV}), the dependence of brightness on high voltages remains to be studied. In the meantime, for ease of comparison, we maintain the original definition of brightness \cite{Borries1939,Sciaini2019,Sudraud1978} rather than using reduced brightness $B_{r}=B/V$.

For applications such as focused ion beams, the energy spread of the ion beam $\Delta E_{0}$ is another important parameter. A source-intrinsic narrow energy distribution is always preferable. Typical intrinsic energy-spread values can be found in the literature \cite{Tsong1964}. For LMIS sources, principally used in focused ion microscopes, the order of magnitude is $\Delta E_{0}\approx 10$ eV \cite{Mair1996} while for GFIS sources, such as our CIS, it is much lower: $\Delta E_{0} \approx 1$ eV \cite{Tsong1964}. Although not yet measured in our context, we can reasonably expect the CIS intrinsic energy-spread $\Delta E_{0}$ to be smaller than for LMIS. Moreover, the low residual pressure inherent to needle-in-capillary type GFIS is an advantage in terms of avoiding collision energy-spread. \\
The order of magnitude of the brightness given above, $B\approx 3 \times 10^{11}$ A/m$^2$/sr, is very promising, although the two key parameters, brightness and energy spread, have yet to be fully characterised. Such a characterisation is beyond the scope of this work and is planned for later, when the tip is mounted in an ion column.

\section{CONCLUSION}
We investigated the gas flow regimes of a coaxial ion source under a range of pressures. Flow conductance measurements taking into account the viscosity and molecular mass of the different gases consistently exhibit a generic pattern: three different flow regimes are observed with increasing upstream pressure, resulting in an S-shaped curve. Even when the geometry of the structure is different, the pattern remains consistent with the new geometrical parameters (see Appendix \ref{CvsGeom} for more information).

The ionisation current was measured here with respect to pressure and extraction voltage. Current is proportional to pressure in the low-pressure chamber and increases with voltage according to a power law in the regime with an ionisation probability of 100$\%$. Similar experiments in a partial pressure configuration show that the CIS geometry enables local pressure at the tip to reach values at least two orders of magnitude higher than the pressure in the chamber. Allowing much higher currents for the same pressure is therefore an advantage of using the CIS as a source for ion optics. At the same time, it minimises the impediments to beam propagation, known to increase energy spread. Optimising the source geometry may even increase this gain. Moreover, a promising estimation of brightness is obtained for our CIS, which compares favourably to the brightness reported for liquid metal ion sources. 
Further characterisations will obviously follow, but, should its brightness be confirmed and its stability be demonstrated to be sufficient for practical applications, this room temperature CIS provides a valuable option as a simple, high-brightness, monochromatic noble gas ion source.

\begin{acknowledgments}
The support of Orsay Physics Company and the Region Sud, providing a joint grant for one of the authors (D.B) and financial support, is gratefully acknowledged. The authors would like to thank Marjorie Sweetko for improving the English of this article, Hubert Klein for his help with the graphics, and Stéphane Varta, from CEA Cadarache, for lending the mass spectrometer.
\end{acknowledgments}

\section*{Data Availability Statement}

The data that support the findings of this study are available from the corresponding author upon reasonable request.

\appendix
\nocite{*}
\bibliography{biblio}

\section{Effect of CIS geometrical parameters on conductance}
\label{CvsGeom}
Pressure measurements $P_{lp}$ vs. $P_{hp}$ were performed for 5 different gases and two different types of CIS structure. The length of the stainless-steel capillary is $L=6$mm and the diameter of the tungsten wire is $d = 125 \mu$m for each CIS structure. The diameter of the stainless-steel capillary changes, giving an area of the opened section: $S=\pi \frac{D+d}{2}\times\frac{D-d}{2}=a\times b$ (see Figure \ref{coax}). For the CIS structure named $s_7$, $D = 150 \mu$m, we then have $S_{7}=a_{7}\times b_{7}=432\times12.5 \mu$m$^2$; and for the CIS structure $s_8$, presented in the article, $D = 170 \mu$m, $S_{8}=a_{8}\times b_{8}=463\times22.5 \mu$m$^2$.

\begin{figure*}
\includegraphics{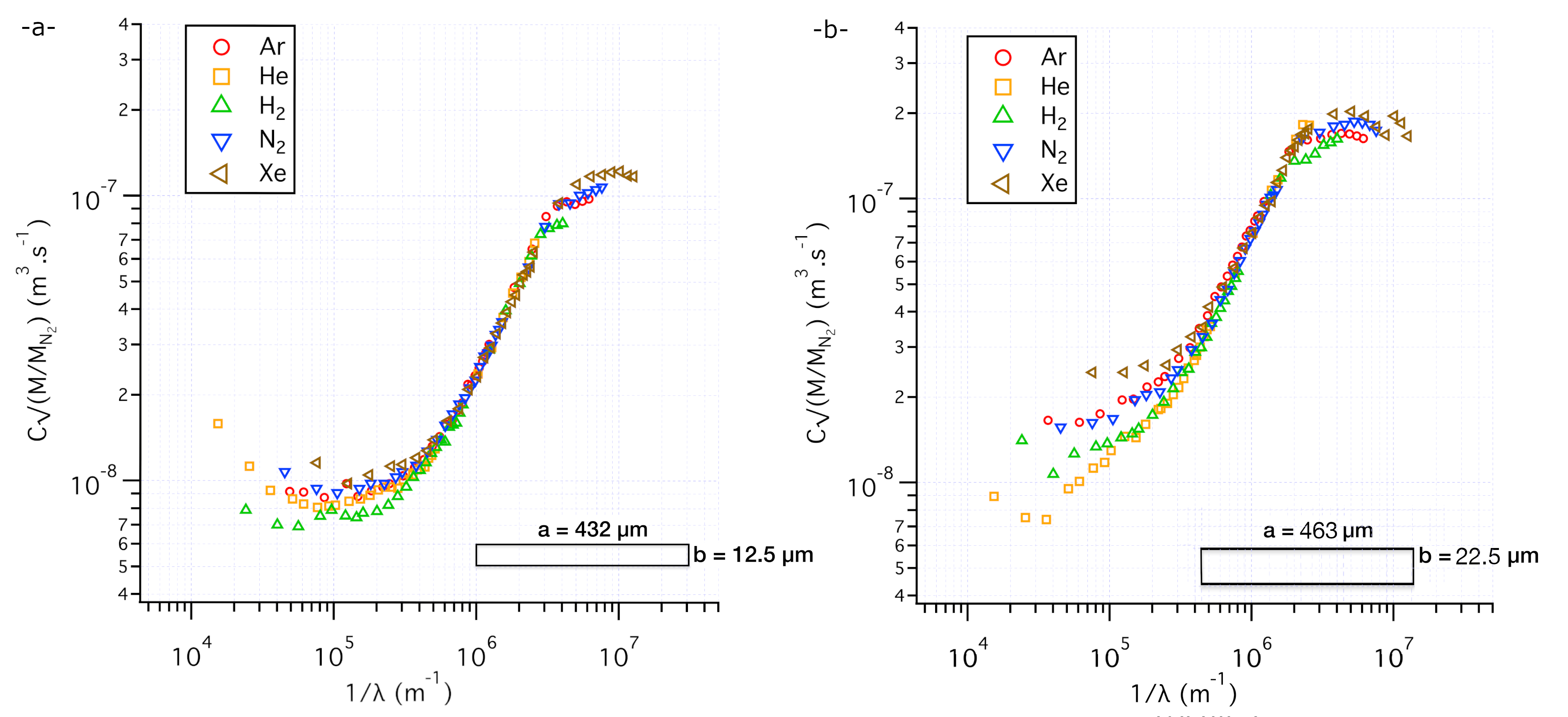}
\caption{\label{ConducGeom} Reduced conductance $C^{*}$ vs. $1/\lambda$ (the inverse of the mean free path of the high-pressure chamber) for 5 gases.-a- For CIS geometry $S_{7}=a_{7}\times b_{7}=432\times12.5 \mu$m$^2$; -b- For CIS geometry $S_{8}=a_{8}\times b_{8}=463\times22.5 \mu$m$^2$.}
\end{figure*}

The reduced conductance versus the inverse of the mean free path of the high-pressure chamber is plotted in Figure \ref{ConducGeom} for each gas and each structure. The generic S-shaped curve described in Section \ref{conductance} appears for both experiments.\\
Ratio between experimental conductances in the choked-flow regime of the two structures gives: $0.53\pm{0.08}$. Using Equation \ref{eq:chock} of the conductance for a rectangular aperture in the choked-flow regime, this ratio can be calculated: $a_{7}b_{7}/(a_{8}b_{8}) \approx 0.52$. This confirms that the value of $C_0$ remains constant. From our data, we derive $C_{0}\approx0.1$, which is close to the expected value \cite{Descoins2008a,santeler1986}.\\
\begin{table}
\caption{\label{beff} The inverted mean free path at the viscous-to-choked flow conductance cross-over ($1/\lambda_{co})$ for structures $s_7$ and $s_8$; the width $b$ of the structure; $(b^{2}/\lambda_{co})$; and, the effective characteristic length ($b_{eff}$) calculated with Equation \ref{eq:beff}.}
\begin{ruledtabular}
\begin{tabular}{c|c|c|c|c}
structure & $1/\lambda_{co}$ (m$^{-1}$) & b ($\mu$m) & $b^{2}/\lambda_{co}$ ($\mu$m) & $b_{eff}$ ($\mu$m) \\
\hline
$s_7$ & $(4\pm1)\times 10^{6}$  & $12.5$ & $625$ &  $(45\pm8)$ \\
\hline
$s_8$ & $(2.5\pm1.0)\times 10^{6}$ & $22.5$ & $1265$ & $(57\pm 9)$ \\
\end{tabular}
\end{ruledtabular}
\end{table}
We take our consistency tests farther, looking at different regimes. We now consider the transition between the choked-flow and the viscous-flow regimes. From Equations \ref{eq:visc} and \ref{eq:chock}, the crossover obtained for $(1/\lambda_{co})$ should correspond to the same $(b^2/\lambda_{co})$ value for the two structures. The observed transition from viscous- to choked-flow occurs for structure $s_7$ at $(1/\lambda_{co})_{s7}=(4\pm1)\times 10^{6}$m$^{-1}$ and for structure $s_8$ at $(1/\lambda_{co})_{s8}=(2.5\pm1.0)\times 10^{6}$m$^{-1}$. However, calculating $(b^2/\lambda_{co})$, we obtain significantly different values, as can be seen in Table \ref{beff}. This suggests our geometry is not appropriately described by a perfect annulus approximated by a rectangular aperture of width $b$. The tungsten wire can lie along the inner stainless steel capillary tube. In such a geometry, the characteristic length of the conductance channel, which we will call $b_{eff}$, will be larger than $b$. $b_{eff}$ is the appropriate length scale for comparison with the mean free path. While we have seen above that the opening area is a good, constant parameter, if we stick with the simple equations of the rectangular geometry, $a_{eff}$ will be somewhat smaller than $a$. We thus calculate $b_{eff}$ by equalising the expression of conductance in the viscous-flow regime (see Equation \ref{eq:visc}) and in the choked-flow regime (see Equation \ref{eq:chock}). We obtain:
\begin{equation}
    (\frac{b_{eff}^2}{\lambda_{co}})=48\sqrt{\frac{1}{2\pi}}C_{0}\Gamma L \label{eq:beff}
\end{equation}

We note that the values of $b_{eff}$, reported in Table \ref{beff} exceed $2 \times b$, which would be the upper limit in a simple geometrical approach. However $b_{eff}$ is the appropriate parameter to calculate the so-called Knudsen number $K=\lambda/b_{eff}$ commonly used in the literature \cite{Knudsen,steckelmacher1966,Steckelmacher1986,vacuum, hanlon}.

\section{Emission angle of the ion beam}
\label{emissionangle}
As described in Section \ref{MaterialsandMethods}, the CIS structure is mounted in an ion projection microscope. This experimental setup can be used to measure the size of the virtual source. The CIS structure is facing the extractor grid, which can be approached to increase the magnification of the projection image formed on a 4cm diameter fluorescent screen 40cm away. To perform this experiment, the CIS structure is positively electrically supplied, and the extractor is electrically grounded. The entrance of the MCP/fluorescent screen assembly is also grounded so that the ion beam is field-free throughout. Then, to maintain a constant electric field at the tip when the extractor is approached, the voltage is decreased to keep the ion emission current constant.
Both the CIS structure and the extractor grid are installed on the same rotary manipulator in front of the screen to allow the entire emission beam to be probed, since projection size limits the detection angle to $5^{\circ}$.

\begin{figure*}
\includegraphics{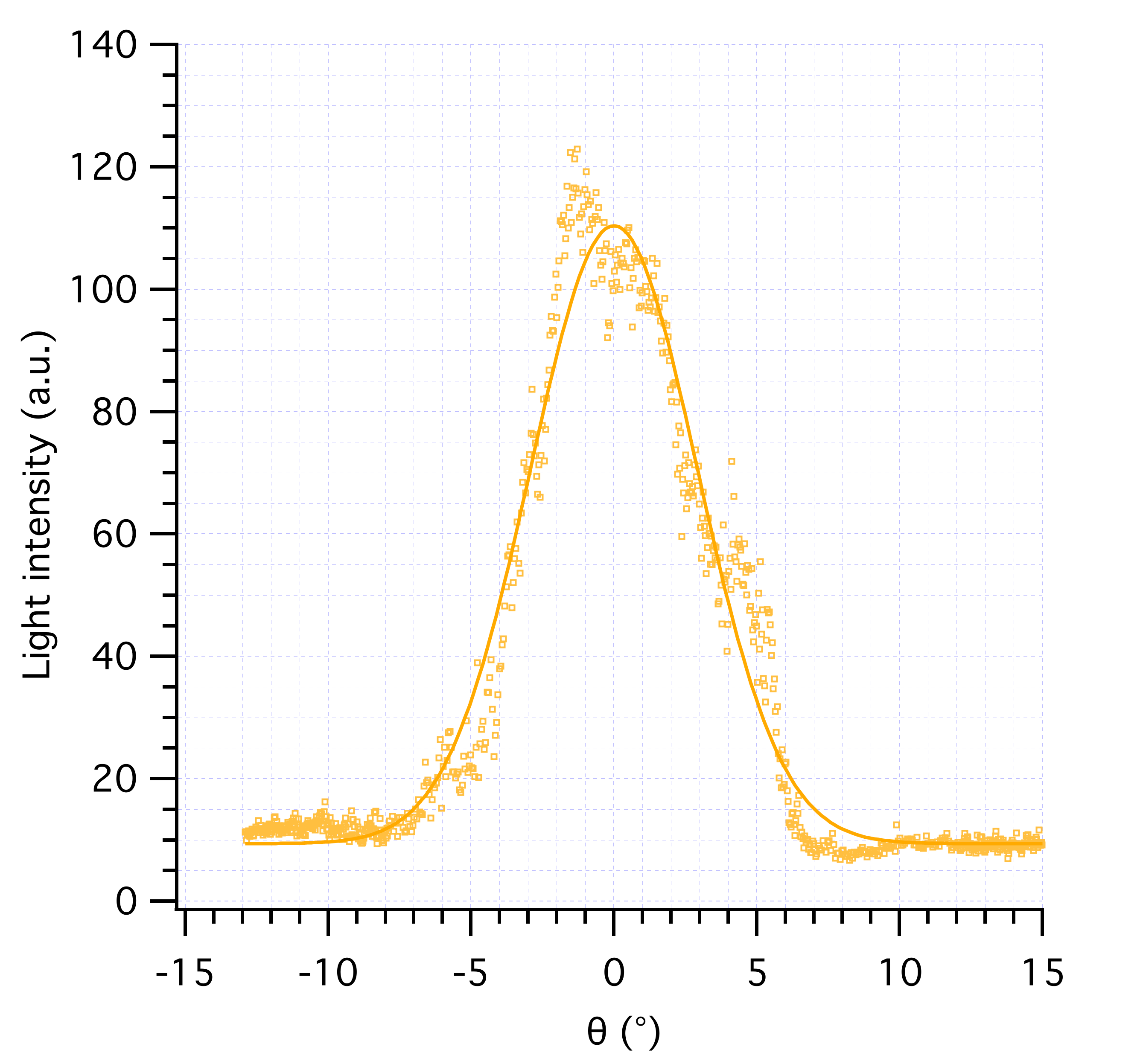}
\caption{\label{angle} Divergence angle of the beam. The mean light intensity is recorded on the 4cm screen while the structure is rotated at an angle $\theta$.}
\end{figure*}

To observe the entire emission beam, we turn the structure, recording the image on the screen. The recorded light intensity on the image is proportional to the ion intensity detected on the screen. Thus, the light intensity value can be plotted versus the rotation angle. It is presented in Figure \ref{angle} for a CIS structure supplied at $V=12$kV and a pressure in the low-pressure chamber of about $P_{lp}=2.7\times 10^{-3}$Pa. This angle gives $\Omega=0.024$sr, which is the solid angle used herein for brightness estimation.

\section{Effective pumping speed}
\label{effpumpsped}
\begin{figure*}
\includegraphics{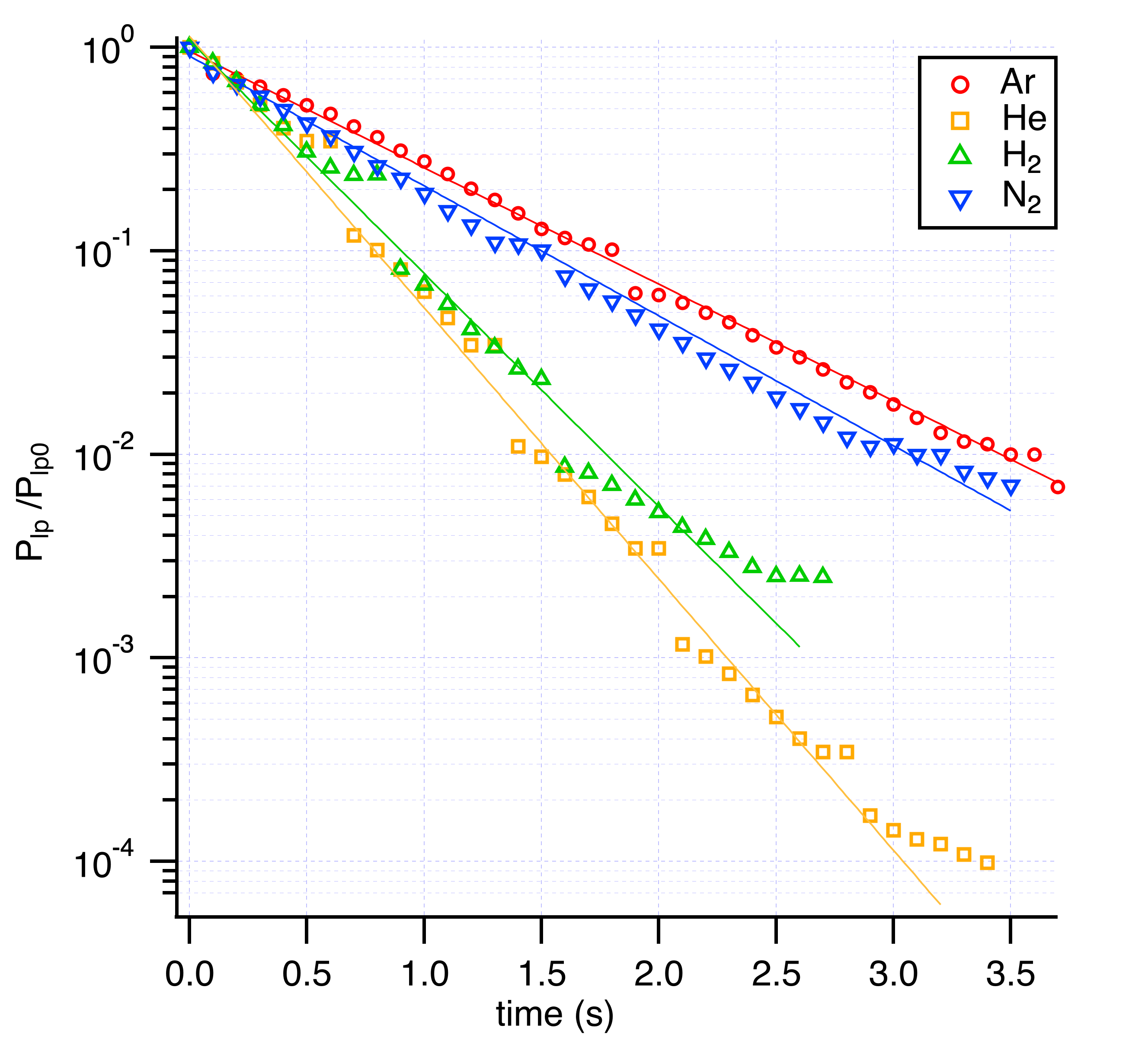}
\caption{\label{pumpspeed} Effective pumping speed measurement. The (exponential) decrease in pressure $P_{lp}$ in the low-pressure chamber is recorded over time $t$ using the mass spectrometer.}
\end{figure*}

To complement the effective pumping speed measurements given in Section \ref{Results}, Figure \ref{pumpspeed} shows the pressure decrease in the low-pressure chamber $P_{lp}/P_{lp0}$ versus $time$ recorded with the mass spectrometer for 4 gases. For easier comparison of the four curves, pressure $P_{lp}$ is divided by the highest pressure $P_{lp0}$ measured at $t=0$s for each gas ($P_{lp0}\approx 10^{-3}$Pa).
Recalling that  $P_{lp}/P_{lp0}=\exp(-S_{e}t/V)$, to find the effective pumping speed, we need to know the volume $V$ of the low-pressure chamber. Here, we have $V=(67\pm5)$L.
The effective pumping speeds found are presented in Table \ref{speed} and were used in the treatment of data presented in Appendix \ref{CvsGeom}. For two different structures and five different gases, these values permit a consistent analysis.

\end{document}